\documentclass[aps,prb,twocolumn,superscriptaddress]{revtex4-2}

\usepackage{amssymb}
\usepackage{amsmath}
\usepackage{graphicx}
\usepackage{array}
\usepackage{multirow}
\usepackage{setspace}

\usepackage[colorlinks,linkcolor=blue,anchorcolor=blue,citecolor=red]{hyperref} 

\usepackage{booktabs}
\usepackage{diagbox}

\begin{document}

\title{Local magnetic structure in fully and partially ordered V$_2$$X$Al Heusler alloys ($X$=Cr, Mn, Fe, Co, Ni)}
\author{Zhenyang Xie}
\affiliation{Institute for Applied Physics and Department of Physics, University of Science and Technology Beijing, Beijing 100083, China}
\author{Jitong Song}
\affiliation{Institute for Applied Physics and Department of Physics, University of Science and Technology Beijing, Beijing 100083, China}
\author{Yuntao Wu}
\affiliation{Institute for Applied Physics and Department of Physics, University of Science and Technology Beijing, Beijing 100083, China}
\author{Yuanji Xu}
\email{yuanjixu@ustb.edu.cn}
\affiliation{Institute for Applied Physics and Department of Physics, University of Science and Technology Beijing, Beijing 100083, China}
\author{Fuyang Tian}
\email{fuyang@ustb.edu.cn}
\affiliation{Institute for Applied Physics and Department of Physics, University of Science and Technology Beijing, Beijing 100083, China}
\affiliation{State Key Laboratory for Advanced Metals and Materials, University of Science and Technology Beijing, Beijing, 100083, China}

\date{\today}

\begin{abstract}
Multicomponent Heusler alloys exhibit various magnetic properties arising from their diverse atomic compositions and crystal structures. Identifying the general physical principles that govern these behaviors is essential for advancing their potential in spintronic applications. In this work, we combine density functional theory with atomistic Monte Carlo simulations to investigate the magnetic ground states, finite-temperature magnetic transitions, and electronic structures of fully-ordered $L2_1$-, $XA$-type, and partially-ordered V$_2X$Al ($X=$ Cr, Mn, Fe, Co, Ni) Heusler alloys. We propose the concept of magnetic motifs, defined as V-$X$-V triangular pathway connected by the nearest-neighbor (NN) exchange interactions $J_{\mathrm{V-}X}$. Within this framework, the magnetic ground states and transition temperatures across the V$_2X$Al family can be consistently understood. The magnetic order is primarily governed by the NN $J_{\mathrm{V-}X}$ interactions in the triangular motifs, while the transition temperatures are additionally influenced by $J_{X-X}$ couplings. Furthermore, the magnetic motifs are still proven to be effective in our calculations on partially-ordered V$_2$$X$Al alloys from $L2_1$ to $XA$-type structures. Our results suggest that the concept of magnetic motifs provides a unifying principle for understanding magnetic ordering in V-based Heusler alloys and could serve as a candidate guide for exploring magnetism and designing advanced spintronic materials in a broader class of Heusler systems.
\end{abstract}

\maketitle

\section{INTRODUCTION}

Heusler alloys, long regarded as prototypical spintronic materials owing to their half-metallic nature, are capable of delivering nearly $100\%$ spin polarization at and even well above room temperature \cite{Gupta2023,Venkateswara2023,Nishihaya2025,deGroot1983,Zutic2004,Katsnelson2008}. Consequently, the discovery and design of high-performance Heusler compounds remain a central objective of both experimental and theoretical research \cite{Graf2011}. On the experimental side, typical full-Heusler alloys such as Co$_2$CrAl and Co$_2$TiSn have been shown to sustain ferromagnetic order up to room temperature, with Curie temperatures $T_{\mathrm{C}}$ of approximately 335 K and 333 K, respectively \cite{Husmann2006_PRB,Shigeta2018_PRB}. Over the past two decades, theoretical efforts, particularly high-throughput first-principles calculations combined with Monte Carlo or spin-model simulations, have identified dozens of full-Heusler systems exhibiting magnetic anisotropy energies exceeding $1~\mathrm{MJ/m^{3}}$ together with Curie temperatures above $500~\mathrm{K}$ \cite{Kubler2007_PRB,Fortunato2024}. These advances have substantially expanded the library of high-performance magnets and promising high-$T_{\mathrm{C}}$ candidates for spintronic applications \cite{Marathe2023}.

In particular, full-Heusler alloys with the general formula $X_2YZ$ ($Y=$ transition metals) have attracted considerable attention owing to their rich magnetic properties, shape-memory behavior, and sensitivity to chemical disorder \cite{Guo2025,Picozzi2004,Pons2018_PRB}. For instance, Co$_2$VAl exhibits ferromagnetic ordering with a Curie temperature close to room temperature, while its $T_\mathrm{C}$ decreases under applied external pressure \cite{Kanomata2010_PRB}. In contrast, Mn$_2$VGa shows ferrimagnetic ordering with a significantly higher Curie temperature of 784 K \cite{Suto2024}. The inverse-Heusler ($XA$-type) compound Mn$_2$CoAl also displays robust ferromagnetism, with a $T{_\mathrm{C}}$ of 720 K at ambient pressure \cite{Ouardi2013_PRL}, whereas Mn$_2$FeSi, despite sharing the same crystal structure, has an anomalously low $T_{\mathrm{C}}$ of only 67 K \cite{Zivotsky2022_Mn2FeSi}. Moreover, both theoretical and experimental studies of $L2_1$-type full-Heusler alloys have revealed complex magnetic behavior. For example, Co$_2$FeSi exhibits a remarkably high Curie temperature of about 1100 K \cite{Wurmehl2005_PRB}, while Fe$_2$VAl remains paramagnetic up to room temperature. Such diverse magnetic orders are generally attributed to variations in stoichiometry and chemical disorder \cite{Singh1998}, raising fundamental questions about the key factors that govern the magnetic properties of Heusler systems.

Recently, chemical disorder has attracted growing attention as an effective means of tuning magnetic order. Disorder can substantially modify magnetic exchange interactions, influence phase stability, and reshape the overall magnetic state of a material \cite{Khmelevskyi2015_PRB,Sokolovskiy2012_PRB,Decolvenaere2019_PRM}. For example, studies on Ru$_2$Mn$Z$ ($Z=$ Sn, Sb, Ge, Si) full-Heusler alloys have demonstrated the strong impact of chemical disorder on their magnetic behavior \cite{Khmelevskyi2015_PRB}, with Co$_2$FeSn exhibiting enhanced magnetism and pronounced size-dependent disorder effects \cite{Li2013_MNL}. Similarly, Smith \textit{et al.} recently employed DC magnetron sputtering to grow epitaxial V$_2$FeAl thin films \cite{Smith2023}. Although V$_2$FeAl shows a relatively high Curie temperature experimentally, its crystal structure appears to be a mixture of the $XA$- and $L2_1$-type phases, complicated further by intrinsic disorder \cite{Smith2023}. Other works have demonstrated that increasing the degree of chemical disorder leads to a pronounced suppression of spontaneous magnetization and Curie temperature \cite{Smith2023,Miroshkina2022_PRB1}. From a broader physical perspective, the deliberate introduction of chemical disorder not only provides an alternative strategy for tuning magnetism but also establishes a promising platform to investigate the fundamental mechanisms underlying complex magnetic orders in Heusler alloys \cite{Decolvenaere2019_PRM}.

Although the magnetic properties of full-Heusler alloys have been extensively investigated experimentally, the underlying mechanisms remain elusive due to their complex atomic compositions, multi-sublattice character, and the presence of chemical disorder. Using first-principles approaches, previous works have systematically evaluated the magnetic exchange couplings $J_{ij}$ for several Heusler alloy families employing frozen-magnon and linear-response techniques \cite{Galanakis2002,Chico2016}. For vanadium-based antiferromagnets, existing studies have primarily concentrated on the ordered $L2_1$-type structure of V$_2Y$Al ($Y=$ Nb, Ta) \cite{Kuroda2020}. Follow-up investigations extended this work by comparing the $L2_1$ phase with the inverse $XA$ phase across a broader range of compounds. In particular, Kuroda \textit{et al.} reported that $L2_1$-type V$_2$NbAl and V$_2$TaAl favor antiferromagnetic coupling, whereas in V$_2Y$Si ($Y=$ Ti, Zr, Hf), the $XA$ structure often competes with or destabilizes the $L2_1$ phase \cite{Kuroda2020}. Despite these advances, a synthetically theoretical investigation of magentism of V element in V$_2X$Al compounds remains limited, especially when $X$ is considered as a magnetic element. A comprehensive microscopic understanding of the mechanisms that govern magnetism in both fully- and partially-ordered V-based full-Heusler alloys is therefore essential.

In this work, we investigate the ground-state magnetic orders and electronic structures of V$_2X$Al ($X=$ Cr, Mn, Fe, Co, Ni) Heusler alloys using density functional theory. V atoms play a key role in regulating the magnetic moments of the five magnetic transition elements, influencing both the orientation and magnitude of the magnetic moments. For the fully-ordered $L2_1$-type V$_2X$Al, V$_2$CrAl and V$_2$MnAl exhibit ferrimagnetic order, whereas the remaining compounds adopt ferromagnetic order. The situation becomes more intricate in the inverse $XA$-type V$_2X$Al. While V$_2$CrAl displays a nearly vanishing net magnetic moment with ferromagnetic order between the two V atoms on different Wyckoff sites, the other four alloys exhibit ferrimagnetic order with antiparallel spin alignment between the two V sublattices. Despite this diversity, the magnetic orders across V$_2X$Al compounds can be consistently interpreted in terms of the V-$X$-V triangular pathway, governed by the NN magnetic exchange coupling $J_{\mathrm{V-}X}$. Furthermore, finite-temperature magnetic transitions were systematically investigated through atomistic Monte Carlo simulations, which reveal that the sequence of Curie temperatures among different compositions and lattice types is primarily determined by $J_{\mathrm{V-}X}$, with additional contributions from $J_{X-X}$ exchange interactions. Importantly, we find that the V-$X$-V triangular pathway framework remains valid even in partially-ordered systems from $L2_1$ to $XA$-type structures. Building on this insight, we propose the concept of ``magnetic motifs'', defined as the V-$X$-V triangular motif connected by NN exchange interactions $J_{\mathrm{V-}X}$. We recognize that the magnetic motifs could provide a promising concept for exploring the magnetic properties in more Heusler alloys.

\begin{figure}
\begin{center}
\includegraphics[width=0.45\textwidth]{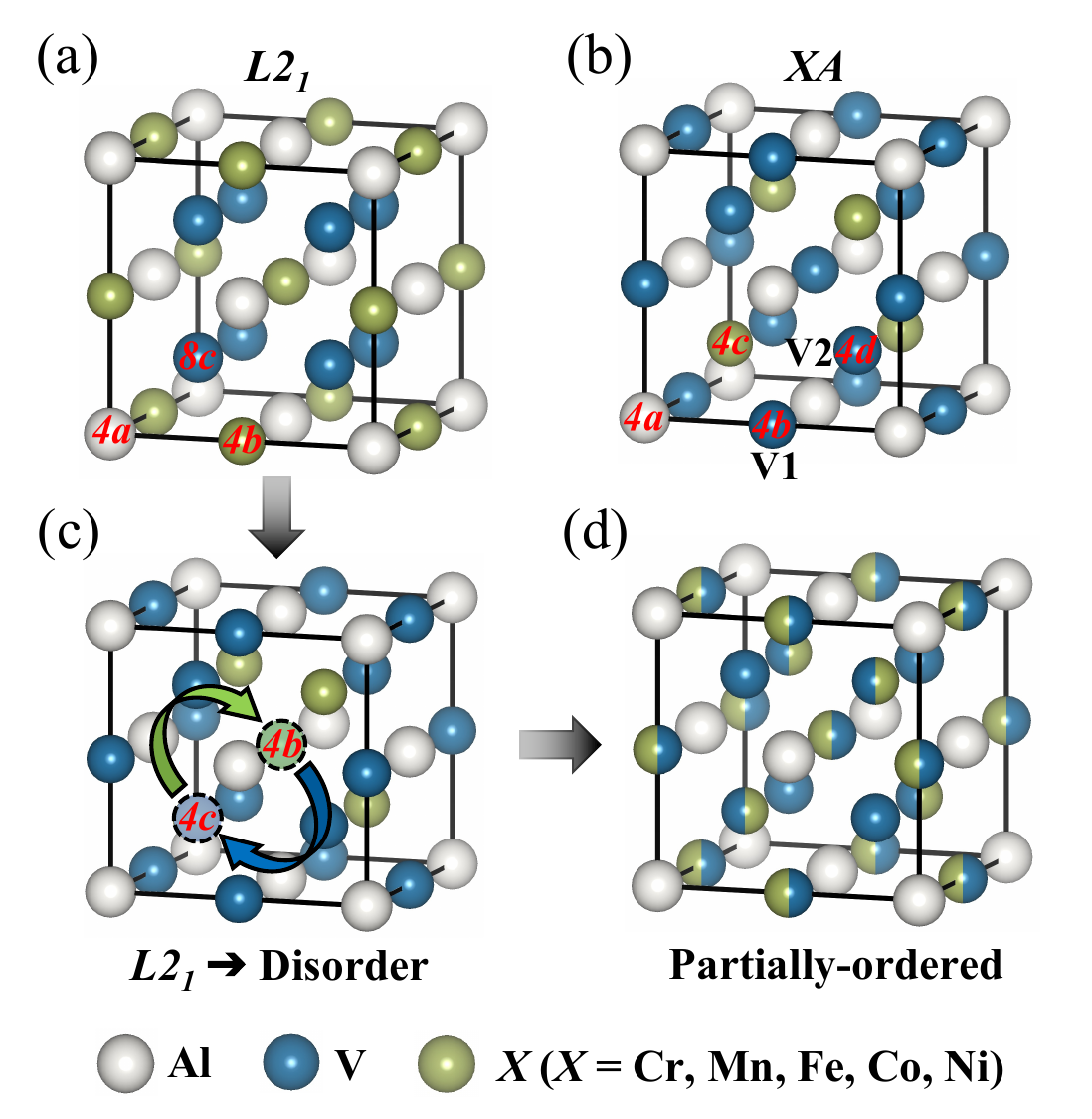}
\caption{Crystal structures of V$_2X$Al ($X=$ Cr, Mn, Fe, Co, Ni) Heusler alloys. (a) Fully-ordered $L2_1$-type full-Heusler ($Fm\bar{3}m$). (b) Fully-ordered $XA$-type inverse-Heusler ($F\bar{4}3m$). (c) Schematic pathway from $L2_1$ to $XA$ through disorder. (d) Partially-ordered V$_2X$Al structure considered in this work.}
\label{fig1}
\end{center}
\end{figure}

\section{METHODS}

All calculations were performed based on spin-polarized density functional theory \cite{Hohenberg1964,Kohn1965}. To get the equilibrium-state structures of Heusler alloys, we used the VASP code to relax the crystal structures  \cite{Kresse1996,Kresse1999}, which implements the projector-augmented wave (PAW) method \cite{Blochl1994PAW}, and the generalized gradient approximation in the form of Perdew-Burke-Ernzerhof (GGA-PBE) functional was employed to represent the exchange-correlation potential with a dense $k$-mesh of 10×10×10. The plane-wave cutoff was chosen to be 425 eV. The convergence criteria for the total energy and ionic forces were set to $10^{-7}\,\mathrm{eV}$ and $10^{-3}\,\mathrm{eV/\AA}$, respectively.For each composition of V$_2X$Al, we first obtained the relaxed lattice constants of the fully-ordered $L2_1$ structure. Based on these optimized configurations, partially-ordered structures were constructed by exchanging V and $X$ atoms, as illustrated in Fig.~\ref{fig1}(c). Each partially ordered model was subsequently subjected to an independent lattice relaxation. In this study, all results are obtained from the standard DFT calculations without explicitly accounting for strong electronic correlations. Previous DFT investigations have shown that exchange interactions and Curie temperatures in many Heusler alloys can be accurately reproduced without the inclusion of a Hubbard $U$ term, yielding good agreement with experimental measurements \cite{Kubler2007,Sasioglu2005,Pajda2001}. For comparison, the results of DFT+$U$ calculations are provided in the Supplementary Information \cite{Supp} (see also references \cite{Liu2023,Skaftouros2013_PRB,Zhang2012} therein). While the inclusion of the Coulomb interaction $U$ does not change the main physical trends reported here, the DFT+$U$ approach generally overestimates the magnitude of the exchange interactions, leading to unrealistically high Curie temperatures.

We calculated the magnetic orders and electronic structures by using the \textsc{Atomic-orbital Based Ab-initio Computation at USTC} (ABACUS) package \cite{Li2016,Lin2025}, an open-source first-principles platform with advanced ground- and excited-state modules, and interfaces to Wannier90. We employed the SG15 optimized norm-conserving Vanderbilt (ONCV) pseudopotentials \cite{Schlipf2015} and the same exchange correlation functional GGA-PBE \cite{Perdew1996PBE}. The Kohn–Sham electron wave functions were expanded in terms of a linear combination of atomic orbitals (LCAO) associated with the SG15 ONCV pseudopotentials (2.0) \cite{Lin2025}. After convergence tests, the grid cutoff for numerical integrations was set to 100 Ry.

To explore the magnetic properties in V$_2$$X$Al, we employed the TB2J package \cite{He2021}, which is based on the use of Green’s functions. The real-space Heisenberg exchange parameters
\[
  J_{ij} = \frac{1}{4\pi} \int_{-\infty}^{E_{\text F}}
           \mathrm dE\,
           \operatorname{Im}\!
           \operatorname{Tr}\!\bigl[
              \Delta_i\,G_{ij}^{\uparrow}(E)\,
              \Delta_j\,G_{ji}^{\downarrow}(E)
           \bigr],
\]
where $\Delta_i$ is the on-site exchange splitting and $G_{ij}^{\sigma}(E)$ the spin-resolved Green function. The complete set $\{J_{ij}\}$ defines a classical Heisenberg Hamiltonian
\[
  \mathcal{H} = -\sum_{i\neq j} J_{ij}\,\mathbf S_i \!\cdot\! \mathbf S_j ,
\]
which was solved with the \textsc{Vampire} atomistic spin-dynamics code \cite{Skubic2008}.  

In Monte Carlo simulations, we examined the convergence with respect to both the supercell size and $k$-point sampling. We found that the magnetic transition temperature exhibited negligible variation for supercell sizes ranging from 10 to 20 nm. Similarly, for $k$-point grids ranging from $4 \times 4 \times 4$ to $17 \times 17 \times 17$, the results were essentially unchanged for grids finer than $4 \times 4 \times 4$. Based on these convergence tests, we selected the Monte Carlo supercell of 15 nm and the $k$-point mesh of $9 \times 9 \times 9$ for all subsequent calculations. The simulated system of V$_2$$X$Al is a cubic box of edge length 15 nm, generated by replicating the relaxed face-centered unit cell and containing about $2.1\times10^{5}$ spins, which was evolved with a stochastic spin integrator. Temperatures from $0$ to $1000\ \text{K}$ in $5\, \text{K}$ increments were sampled using $2.5\times10^{3}$ equilibration steps. The Curie temperature $T_{\mathrm C}$ was determined from the inflection point of the magnetization $M(T)$.

\begin{figure}
\begin{center}
\includegraphics[width=0.40\textwidth]{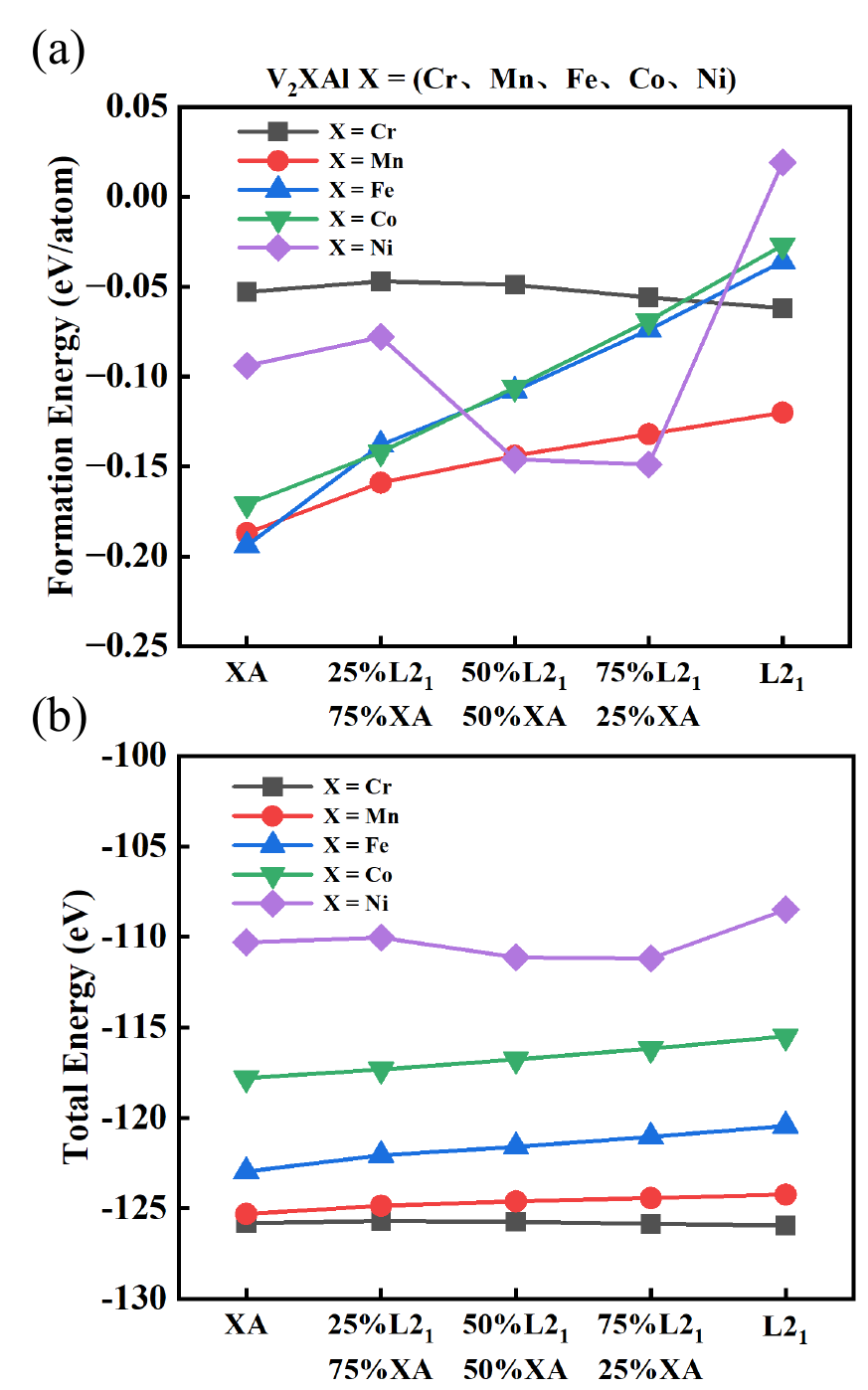}
\caption{(a) Formation energies of V$_2X$Al ($X=$ Cr, Mn, Fe, Co, Ni) Heusler alloys for the fully-ordered ($XA$, $L2_1$) and partially-ordered structural configurations. (b) Total energies of V$_2X$Al for the corresponding fully-ordered and partially-ordered structures.}
\label{fig2}
\end{center}
\end{figure}

\begin{table*}[htbp]
  \caption{Spin magnetic moments $m$ of each inequivalent atomic site and the total magnetic moment $m_\mathrm{tot}$ ($\mu_\mathrm{B}$/f.u.) for V$_2X$Al ($X=$ Cr, Mn, Fe, Co, Ni) compounds in the $L2_1$-type structure, as obtained in this work and compared with previous literature \cite{Goraus2023}.}
  \label{tab:L21}
  \centering
  \small
  \setlength{\tabcolsep}{8pt} \renewcommand{\arraystretch}{1.1}
  \begin{tabular}{l r r r r l}
    \toprule
    \textbf{Compound}
      & $m_{\mathrm V}(4b)$ 
      & $m_{X}(8c)$ 
      & $m_{\mathrm Al}(4a)$ 
      & $m_{\mathrm{tot}}$
      & \textbf{Source} \\
    \midrule
    $L2_1$-V$_2$CrAl & 0.901  & -0.742  &  0.026 &  1.085 & This work \\
    \addlinespace[2pt]
    $L2_1$-V$_2$MnAl & 1.802  & -2.018  &  0.068 &  1.653 & This work \\
    $L2_1$-V$_2$MnAl & 1.667  & -1.653  & -0.040 &  1.641 & Jerzy \textit{et al.}\ \cite{Goraus2023} \\
    \addlinespace[2pt]
    $L2_1$-V$_2$FeAl & 0.537  &  1.917  & -0.021 &  2.970 & This work \\
    $L2_1$-V$_2$FeAl & 0.362  &  1.889  & -0.014 &  2.599 & Jerzy \textit{et al.}\ \cite{Goraus2023} \\
    \addlinespace[2pt]
    $L2_1$-V$_2$CoAl & 1.238  &  1.136  &  0.024 &  3.636 & This work \\
    $L2_1$-V$_2$NiAl & 1.249  &  0.265  & -0.005 &  2.757 & This work \\
    \bottomrule
  \end{tabular}
\end{table*}

\section{RESULTS AND DISCUSSION}

\subsection{Structural configuration and thermodynamic stability analysis of V$_2$$X$Al}\label{sec:thermodynamic}

The magnetic ordering in Heusler alloys is strongly influenced by both their atomic composition and crystallographic structure. In this work, we focus on two prototypical lattice types, namely the $L2_1$ and $XA$ structures. As established in previous studies, these two structural variants provide a well-defined platform for examining the role of chemical disorder without altering the overall structural topology \cite{Graf2011,Picozzi2004,Galanakis2002}. As illustrated in Fig.~\ref{fig1}(a), the $L2_1$-type full-Heusler structure (space group $Fm\bar{3}m$, No.225) with the generic formula V$_2X$Al consists of V atoms occupying the 8$c$ $(1/4,1/4,1/4)$ Wyckoff positions, while Al and $X$ atoms are located at the 4$a$ $(0,0,0)$ and 4$b$ $(1/2,1/2,1/2)$ sites, respectively \cite{Graf2011}. By contrast, in the $XA$-type inverse-Heusler structure (space group $F\bar{4}3m$, No.216), shown in Fig.~\ref{fig1}(b), the $X$ atoms occupy the 4$c$ $(1/4,1/4,1/4)$ positions that arise from the symmetry breaking of the 8$c$ sites in the $L2_1$-type structure, while the V atoms are shifted to the 4$b$ sites. From a broader perspective, the redistribution of atomic occupations between the 4$c$ and 4$b$ sites, driven by the evolution from $L2_1$ to $XA$-type structures, highlights the importance of chemical disorder in determining the magnetic properties of these compounds. The details of the relaxed lattice parameters for all configurations are summarized in Supplementary Information. A detailed analysis of disorder effects will be presented in Sec.~\ref{sec:disorder}.

In the V$_2X$Al ($X=$ Cr, Mn, Fe, Co, Ni) Heusler alloys, differences in electronegativity among the constituent atoms play a critical role in determining chemical bonding and, consequently, the structural stability of the compounds. The electronegativity difference between vanadium and the $X$-site elements increases from 0.03 to 0.29 Pauli units as $X$ varies from Cr to Ni, enhancing the strength of the V-$X$ chemical bonds and contributing to improved thermodynamic stability across the series. Similarly, the electronegativity difference between Al and the $X$-site atoms rises from 0.05 to 0.31 Pauli units along the same series, providing additional stabilization through $X$-Al chemical bonding. In contrast, the electronegativity difference between V and Al is relatively small (0.02 Pauli units), lower than that of both the V-$X$ and $X$-Al pairs. Nevertheless, this difference still introduces supplementary bonding that help stabilize the overall Heusler structure.

To assess the stability of the fully ordered and partially ordered V$_2$$X$Al compounds, we calculated both the formation energies and total energies of the possible structural configurations. As shown in Fig.~\ref{fig2}(a), all systems, except for the $L2_1$-type V$_2$NiAl, exhibit negative formation energies, indicating their thermodynamic stability. These results are consistent with the trends reported in the Open Quantum Materials Database (OQMD) \cite{Saal2013,Kirklin2015}. The calculated formation energies further reveal that V$_2$MnAl, V$_2$FeAl, and V$_2$CoAl preferentially adopt the $XA$-type structure, whereas V$_2$CrAl is most stable in the $L2_1$ configuration. For V$_2$NiAl, a partially disordered atomic arrangement is energetically favored. These conclusions align well with previous theoretical and experimental studies on V-based Heusler alloys \cite{Nguyen2024,Zhang2012,Smith2023}. The corresponding total energy results for both the fully-ordered structures and the partially-ordered structural pathways are presented in Fig.~\ref{fig2}(b). Consistent with the formation energy trends, the $XA$-type ordering is the lowest-energy configuration for both V$_2$MnAl and V$_2$FeAl. Moreover, the total energies of the partially-ordered configurations lie between those of the parent $L2_1$- and $XA$-type phases. These findings underscore the importance of considering phase stability as a key criterion in the practical design and optimization of high-performance Heusler-based spintronic devices.

Here, we note that the partially-ordered Heusler alloys are constructed by progressively exchanging V atoms at the 4$b$ sites with $X$ atoms at the 4$c$ sites within a 16-atom cubic cell. This procedure generates a series of partially-ordered ``intermediate” structures. However, we find that different disorder configurations for a given composition do not lead to much variation in the exchange interactions. Therefore, in this study, we present the results for the configuration with the lowest total energy.

\begin{table*}[htbp]
  \caption{Spin magnetic moments $m$ of each inequivalent atomic site and the total magnetic moment $m_\mathrm{tot}$ ($\mu_\mathrm{B}$/f.u.) for V$_2X$Al ($X =$ Cr, Mn, Fe, Co, Ni) compounds in the $XA$-type inverse-Heusler structure. Results from this work are compared with previous studies.}
  \label{tab:XA}
  \centering
  \small
  \setlength{\tabcolsep}{8pt} \renewcommand{\arraystretch}{1.1}
  \begin{tabular}{l r r r r r l}
    \toprule
    \textbf{Compound}
      & $m_{\mathrm V1}(4b)$
      & $m_{\mathrm V2}(4d)$
      & $m_{X}(4c)$
      & $m_{\rm Al}(4a)$
      & $m_{\mathrm{tot}}$
      & \textbf{Source} \\
    \midrule
    $XA$-V$_2$CrAl &  -0.497 &  -1.397 &  1.774 &  -0.026 &  0.145 & This work \\
    \addlinespace[2pt]
    $XA$-V$_2$MnAl &  -0.820 &   1.466 &  1.610 &   0.007 &  2.264 & This work \\
    $XA$-V$_2$MnAl &  -0.689 &   1.440 &  1.340 &  -0.082 &  2.009 & Jerzy \textit{et al.}\ \cite{Goraus2023} \\
    \addlinespace[2pt]
    $XA$-V$_2$FeAl &  -0.447 &   2.043 &  1.179 &   0.002 &  2.777 & This work \\
    $XA$-V$_2$FeAl &  -0.790 &   2.180 &  1.699 &  -0.120 &  2.950 & Jerzy \textit{et al.}\ \cite{Goraus2023} \\
    $XA$-V$_2$FeAl &  -0.310 &   2.110 &  1.20   &   0.090 &  3.090 & Skaftouros \textit{et al.}\ \cite{Skaftouros2013_PRB} \\
    \addlinespace[2pt]
    $XA$-V$_2$CoAl &  -0.404 &   1.801 &  0.556 &   0.043 &  1.996 & This work \\
    $XA$-V$_2$CoAl &  -0.54  &     2.16 &   0.50   &  -0.12   &  2.0     & Zhang \textit{et al.} \ \cite{Zhang2012} \\
    \addlinespace[2pt]
    $XA$-V$_2$NiAl &  -0.526 &   1.527 &  0.107 &   0.039 &  1.148 & This work \\
    $XA$-V$_2$NiAl &  -0.66   &   1.74   &  0.06   &   -0.06 &   1.08 &   Zhang \textit{et al.}\ \cite{Zhang2012} \\
    \bottomrule
  \end{tabular}
\end{table*}

\subsection{Magnetic orders of fully-ordered $L2_1$- and $XA$-V$_2$$X$Al}\label{sec:magneticorder}

We begin by examining the effect of atomic composition on the magnetic order in $L2_1$-type V$_2X$Al ($X=$ Cr, Mn, Fe, Co, Ni), where the crystal structure is relatively simple, containing only three inequivalent Wyckoff sites compared to the more complex $XA$-type structure. As summarized in Table~\ref{tab:L21}, compounds with $X=$ Fe, Co and Ni exhibit ferromagnetic ordering, whereas those with $X=$ Cr and Mn display ferrimagnetic ordering characterized by antiparallel spin alignment between the $X$ and V sublattices. In particular, our calculations show that V$_2$MnAl adopts a ferrimagnetic ground state, with magnetic moments of $+1.802~\mu_\mathrm{B}$ per V atom and $-2.018~\mu_\mathrm{B}$ per Mn atom. By contrast, V$_2$FeAl exhibits ferromagnetic ordering with a total magnetic moment of $2.970~\mu_\mathrm{B}$ per formula unit, in agreement with previous reports \cite{Goraus2023}. Notably, the total magnetic moments of the $X=$ Fe to Ni compounds do not vary monotonically across the series. While the magnetic moments on the V sublattice increase progressively from Fe to Ni, those on the $X$ site decrease rapidly, leading to non-monotonic behavior in the total moment. These complex magnetic orders and the non-monotonic variation in total magnetization are primarily attributed to differences in the underlying electronic structures arising from the atomic composition. A detailed discussion of the interplay between electronic structure and magnetic order will be presented in Sec.~\ref{sec:elecstr}.

To investigate the influence of lattice type on magnetism, we next examine the $XA$-type inverse-Heusler V$_2X$Al compounds, which possess a more complex crystal structure with four inequivalent Wyckoff sites compared to the $L2_1$-type structure. Interestingly, among these systems, only V$_2$CrAl exhibits ferromagnetic coupling between the two inequivalent V sites, while the magnetic moments at the V-4$b$ and $X$-4$c$ sites align antiferromagnetically, as summarized in Tables~\ref{tab:XA}. This contrasts sharply with the comparatively simple magnetic ordering observed in the $L2_1$-type compounds. More specifically, apart from V$_2$CrAl, all other $XA$-type V$_2X$Al compounds ($X$ = Mn, Fe, Co, Ni) display antiparallel alignment of the spin moments between the V atoms at the 4$d$ and 4$b$ sites. Our calculated magnetic moments for $XA$-type V$_2X$Al ($X$ = Mn, Fe, Co) are in good agreement with previous reports \cite{Skaftouros2013_PRB,Goraus2023,Zhang2012}. Furthermore, the half-metallic compounds V$_2$MnAl and V$_2$FeAl exhibit total magnetic moments of 2.26 and 2.78~$\mu_\mathrm{B}$/f.u., respectively, which are close to the ideal values of 2 and 3~$\mu_\mathrm{B}$/f.u. predicted by the generalized Slater–Pauling relation of $M_\mathrm{tot} \approx Z_\mathrm{t}-18$, where $Z_\mathrm{t}$ is the total valence-electron count \cite{Skaftouros2013_PRB,Flowers2017}. Notably, as $X$ varies from Cr to Ni, the local magnetic moment of the $X$ site decreases monotonically (from 1.77 to 0.11~$\mu_\mathrm{B}$), whereas the total moment evolves in a nonmonotonic manner. These results highlight the intricate and composition-dependent magnetic behaviors of inverse-Heusler V$_2X$Al compounds, motivating the need for a more general understanding of their underlying mechanisms.

\begin{figure*}
\begin{center}
\includegraphics[width=0.85\textwidth]{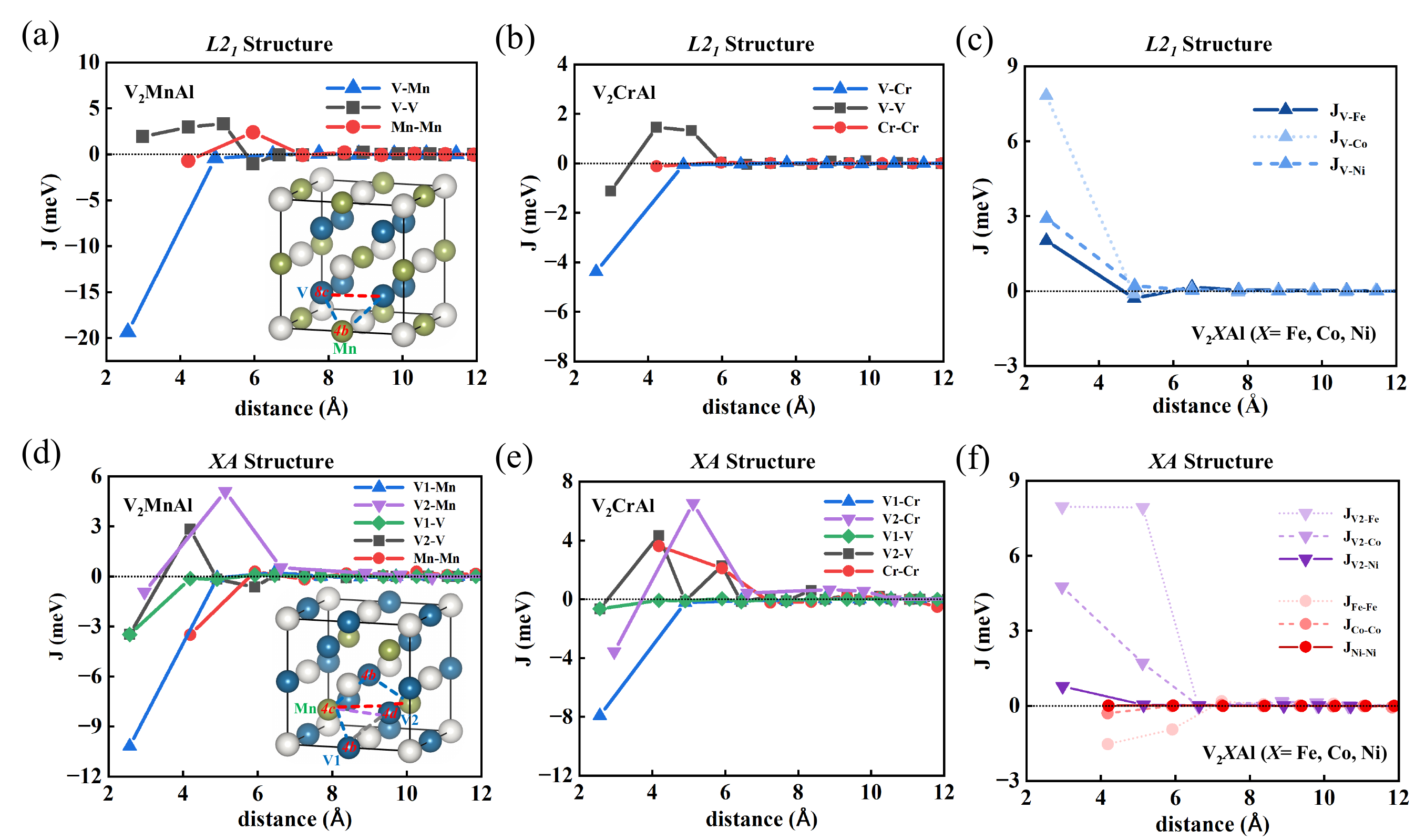}
\caption{Isotropic Heisenberg exchange parameters $J$ for V$_2X$Al compounds. (a, b) $J$ as a function of neighbor distance for $L2_1$-type V$_2$MnAl and V$_2$CrAl, respectively. (c) NN exchange coupling $J_{\mathrm{V-}X}$ in $L2_1$-type compounds with $X=$ Mn, Fe, Co, Ni. (d, e) $J$ as a function of neighbor distance for $XA$-type V$_2$MnAl and V$_2$CrAl, respectively. (f) NN exchange couplings $J_{\mathrm{V2-}X}$ and $J_{X-X}$ in $XA$-type compounds with $X=$ Mn, Fe, Co, Ni. Insets illustrate the triangular exchange pathways responsible for the observed magnetic orders.}
\label{fig3}
\end{center}
\end{figure*}

\subsection{Microscopic magnetic mechanism in $L2_1$- and $XA$-V$_2$$X$Al}

To gain deeper insight into the origin of magnetism in these systems, we investigate the magnetic exchange couplings ($J$) across different atomic compositions and lattice types. As discussed above, in the relatively simple $L2_1$-type V$_2X$Al compounds, only those with $X =$ Cr and Mn exhibit ferrimagnetic order, characterized by anti-parallel spin alignment between the $X$ and V atoms. Correspondingly, as shown in Figs.~\ref{fig3}(a) and \ref{fig3}(b), the NN exchange coupling $J_{\mathrm{V-Mn}}$ ($J_{\mathrm{V-Cr}}$) between Mn (Cr) atoms at the 4$b$ site and V atoms at the 8$c$ site is antiferromagnetic type, which has the largest value among all the other magnetic exchange couplings in these materials. In contrast, for $X =$ Fe, Co and Ni, the NN $J_{\mathrm{V-}X}$ coupling between the $X$ and V atoms is ferromagnetic, as illustrated in Fig.~\ref{fig3}(c). Meanwhile, the exchange interactions between $X$-$X$ atoms are significantly weaker and do not play a decisive role in stabilizing the magnetic order. The V-V interactions, however, exhibit more intricate behavior. In V$_2$MnAl, both the NN and next-nearest-neighbor (NNN) V-V couplings, indicated by the gray line in Fig.~\ref{fig3}(a), are ferromagnetic. However, the NN V-V coupling in V$_2$CrAl, colored by the gray line, acquires a finite antiferromagnetic contribution, as shown in Fig.~\ref{fig3}(b). Taken together, these results indicate that antiferromagnetic V-Mn(Cr)-V exchange pathways form triangular motifs that ultimately stabilize ferromagnetic ordering among the V sublattice. Overall, our exchange-coupling analysis provides a microscopic explanation for the emergence of complex magnetic orders in $L2_1$-type V$_2X$Al compounds.

Here, we investigate the physical mechanisms underlying the magnetic ordering in the more complex $XA$-type V$_2X$Al compounds. In these type compounds, the magnetic moments of V1 (4$b$ site) and $X$ (4$c$ site) are aligned antiparallel, indicating dominant antiferromagnetic exchange coupling. As shown in Figs.~\ref{fig3}(d) and \ref{fig3}(e), the NN exchange couplings $J_{\mathrm{V1-}X}$ (highlighted in blue) are strongly antiferromagnetic ($X=$ Cr, Mn; couplings for $X=$ Fe, Co, Ni are shown in Supplementary Information) and represent the primary factor stabilizing ferrimagnetic ordering between the $X$ and V1 sublattices. In contrast, the NN exchange coupling $J_{\mathrm{V2-}X}$ between $X$ and V2 (4$d$ site), as shown in purple lines in Figs.~\ref{fig3}(e) and \ref{fig3}(f), evolves rapidly from antiferromagnetic to ferromagnetic as $X$ changes from Cr to Ni. This change of $J_{\mathrm{V2-}X}$ is consistent with the corresponding magnetic ordering from ferrimagnetic for $X=$ Cr to ferromagnetic for $X=$ Mn, Fe, Co, and Ni. Notably, although the NN V2-Mn interaction is weakly antiferromagnetic type, the NNN V2-Mn coupling is strongly ferromagnetic, suggesting that V$_2$MnAl lies near a compositional boundary where the magnetic ordering transitions from ferrimagnetic to ferromagnetic with increasing $X$ atomic number.

From our analysis of magnetic orders and exchange couplings, a comprehensive understanding of the overall magnetic behavior in V$_2X$Al across different atomic compositions and lattice types can be established. In $L2_1$-type V$_2X$Al, the diverse magnetic orders can be rationalized in terms of triangular exchange pathways formed by V-$X$-V atoms, as illustrated in the inset of Fig.~\ref{fig3}(a). In $XA$-type V$_2X$Al, the underlying mechanism is similar, but an additional type of magnetic exchange coupling is present. Our results indicate that the V1-$X$-V2 triangular pathway remains a key determinant of magnetic order in $XA$-type compounds, as shown in the inset of Fig.~\ref{fig3}(d). Changes in the type of the $J_{\mathrm{V2-}X}$ coupling substantially influence the magnetic ordering in these materials. Moreover, although the direct NN exchange coupling between $X$-$X$ atoms ($X=$ Mn, Fe, Co, Ni) is antiferromagnetic type in $XA$-V$_2X$Al, the triangular $X$-V1-$X$ pathway ultimately stabilizes ferromagnetic ordering due to the intervening antiferromagnetic X-V1 couplings, as illustrated in the inset of Fig.~\ref{fig3}(d). In summary, despite variations in atomic composition and lattice type, the $X$ site plays a central role in controlling the magnetic properties of V-based V$_2X$Al compounds through triangular pathways formed by NN V and $X$ atoms. We therefore identify these pathway as a candidate ``magnetic motifs''.

\subsection{Magnetic transition temperature of $L2_1$- and $XA$-V$_2$$X$Al}

\begin{figure}
\begin{center}
\includegraphics[width=0.4\textwidth]{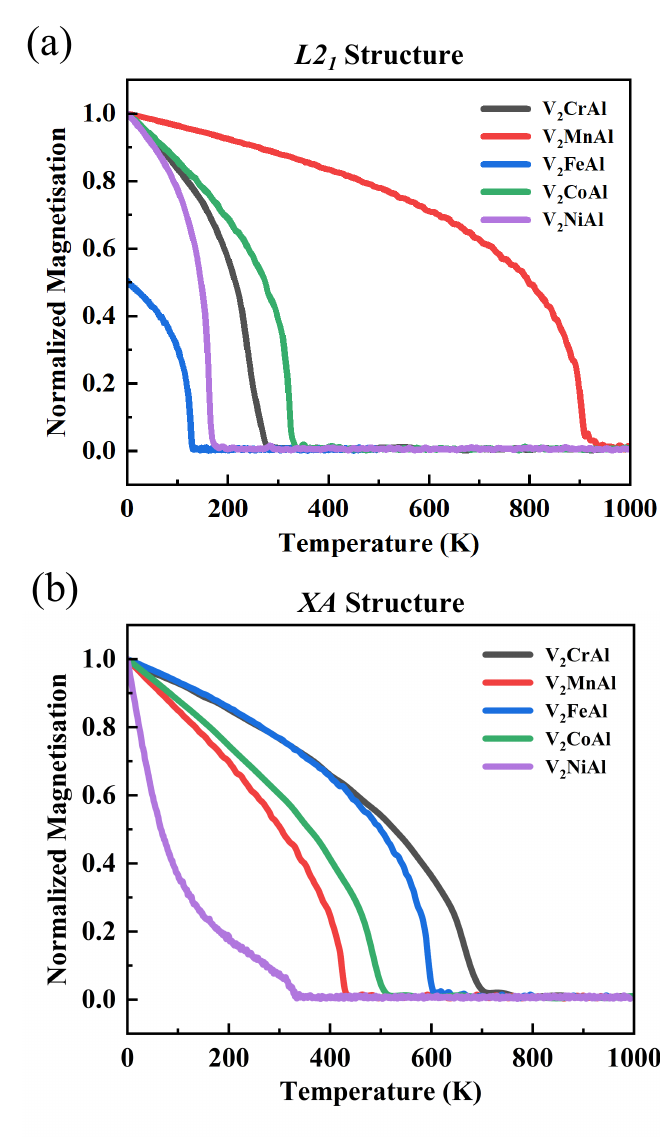}
\caption{(a) Temperature dependence of the normalized magnetic moments for V$_2X$Al compounds in $L2_1$-type V$_2X$Al ($X =$ Cr, Mn, Fe, Co, Ni). (b) Temperature dependence of the normalized magnetic moments for V$_2X$Al compounds in $XA$-type V$_2X$Al.}
\label{fig4}
\end{center}
\end{figure}

The microscopic mechanisms of magnetic exchange couplings directly determine the macroscopic magnetic properties. Based on our calculated exchange couplings in V$_2X$Al, we further simulate the magnetic transition temperatures across this series of Heusler compounds. As shown in Fig.~\ref{fig4}(a), in $L2_1$-type V$_2X$Al, V$_2$MnAl exhibits the highest Curie temperature, reaching approximately 900 K, whereas the Curie temperatures of the remaining compounds ($X =$ Cr, Fe, Co, Ni) are significantly lower, ranging from 100 K to 300 K. The exceptionally high Curie temperature of V$_2$MnAl originates from the effective ferromagnetic coupling between V atoms mediated by the triangular V-Mn-V pathways, which involve strong antiferromagnetic NN couplings $J_{\mathrm{V-Mn}}$ ($\sim -19.4$~meV). This is further reinforced by the direct ferromagnetic V-V interactions. Remarkably, our Monte Carlo simulations reveal that the magnitude of the NN V–$X$ exchange coupling $J_{\mathrm{V-}X}$ predominantly determines the magnetic transition temperature in $L2_1$-type V$_2$$X$Al. Specifically, as shown in Fig.~\ref{fig4}(a), $T_\mathrm{C}$ decreases from 900 K for V$_2$MnAl to 100 K for V$_2$FeAl, consistent with the corresponding NN exchange couplings $J_{\mathrm{V-}X}$, which vary from $-19.4$ to 2.0 meV, as illustrated in Fig.~\ref{fig3}.

The temperature dependence of magnetic phase transitions in $XA$-type V$_2X$Al compounds differs markedly from that in $L2_1$-type structures. As shown in Fig.\ref{fig4}(b), the Curie temperatures in $XA$-type compounds vary within a narrower range, from approximately 700 K to 300 K, compared to the broader variation in $L2_1$-type materials. This narrower range reflects the more complex network of magnetic exchange couplings in the inverse-Heusler structure. Specifically, V$_2$CrAl exhibits the highest $T_\mathrm{C}$ at about 700 K, while the Curie temperature gradually decreases as $X$ changes from Fe to Ni. The magnetic transition temperature of V$_2$MnAl shows a slight anomaly, lying between the values for $X=$ Co and Ni. Despite the apparent complexity of these trends, they can be rationalized using the proposed microscopic mechanism based on the triangular exchange pathways. As discussed above, the NN exchange coupling $J_{\mathrm{V-}X}$ involving the $X$-site dominates the magnetic transition behavior. For instance, the relatively large $J_{\mathrm{V1-Cr}}$ contributes to the higher $T_\mathrm{C}$ in V$_2$CrAl. Although $J_{\mathrm{V1-Mn}}$ is also sizable, the $T_\mathrm{C}$ of V$_2$MnAl is significantly lower due to the additional influence of the $J_{X-X}$ coupling. In V$_2$CrAl, the ferromagnetic $X$–$X$ interactions enhance $T_\mathrm{C}$, whereas in V$_2$MnAl, the $J_{X-X}$ coupling is antiferromagnetic, leading to a notable reduction of $T_\mathrm{C}$. In the series $X=$ Fe, Co, Ni, the gradual decrease of $T_\mathrm{C}$ is primarily driven by the weakening the absolute value of the NN $J_{\mathrm{V1-}X}$ coupling, which diminishes from $3$ to 0.1 meV (see Supplementary Information).

In summary, the magnetic ground states and magnetic transition temperatures of V$_2X$Al compounds can be rationalized within the framework of the V-$X$-V triangular exchange pathway and its associated NN magnetic exchange couplings, applicable to both $L2_1$- and $XA$-type structures. For simplicity, the NN exchange coupling $J_{\mathrm{V-}X}$ primarily governs the magnetic ordering, whereas the combination of $J_{\mathrm{V-}X}$ and $J_{X-X}$ predominantly determines the magnetic transition temperature $T_\mathrm{C}$. Although $T_\mathrm{C}$ depends on both the exchange couplings and the local magnetic moments of the constituent atoms, no straightforward correlation between the magnitude of individual magnetic moments and $T_\mathrm{C}$ is evident. Nevertheless, the observed trends in the dominant exchange couplings provide a practical framework to anticipate and potentially tune the Curie temperatures in these Heusler alloys.

\begin{table*}[htbp]
\caption{Spin magnetic moments $m$ of each inequivalent element and the total magnetic moment $m_\mathrm{tot}$ ($\mu_\mathrm{B}$/f.u.) for partially-ordered V$_2$MnAl alloys considered in this work.}
\label{tab:Mn_XA_L21}
\centering
\small
\setlength{\tabcolsep}{12pt} \renewcommand{\arraystretch}{1.1}
\begin{tabular}{l r r r r r r r}
\toprule
\multirow{2}{*}{\begin{tabular}[c]{@{}l@{}}Structural\\ordering\end{tabular}} &
\multicolumn{2}{c}{$m_{\mathrm{Mn}}$} &
\multicolumn{2}{c}{$m_{\mathrm{V_1}}$} &
\multicolumn{1}{c}{$m_{\mathrm{V_{2}}}$} &
\multicolumn{1}{c}{$m_{\mathrm{Al}}$} &
\multicolumn{1}{c}{$m_{\mathrm{tot}}$} \\
\cmidrule(lr){2-3}\cmidrule(lr){4-5}
& ($4b$) & ($4c$) & ($4b$) & ($4c$) & ($4d$) & ($4a$) & f.u. \\
\cmidrule(lr){1-8}
100\%$L2_1$                    & -2.018 & ––– & ––– & 1.802 & 1.802 & 0.068 & 1.653 \\
75\%$L2_1$/25\%$XA$ & -1.863 & 2.936 & -0.872 & 1.184 & 1.432 & 0.033 & 1.472 \\
50\%$L2_1$/50\%$XA$ & -1.976 & 2.731 & -0.926 & 0.630 & 1.474 & 0.016 & 1.719 \\
25\%$L2_1$/75\%$XA$ & -1.593 & 1.582 & -0.595 & 0.111 & 1.338 & 0.028 & 1.679 \\
100\%$XA$                                   & ––– & 1.610 & -0.820 & ––– & 1.466 & 0.007 & 2.264 \\
\bottomrule
\end{tabular}
\end{table*}

\subsection{Effect of chemical disorder on magnetism of partially-ordered V$_2$$X$Al}\label{sec:disorder}

In V$_2X$Al compounds, the NN magnetic exchange coupling $J_{\mathrm{V-}X}$ plays a central role in determining both the magnetic ground states and the magnetic transition temperatures. This raises the question of whether this framework remains valid in the presence of chemical disorder. In this section, we investigate the evolution of magnetic order and magnetic transition temperatures under partial chemical disorder, introduced by progressively transforming the $L2_1$-type structure toward the $XA$-type structure, as illustrated in Fig.~\ref{fig1}(c) and ~\ref{fig1}(d). The introduction of disorder significantly reduces crystal symmetry, complicating the analysis of magnetic interactions. For clarity, we focus on the $L2_1$-type V$_2$MnAl, which exhibits the highest Curie temperature among its sister compounds. Notably, our calculations show that the magnetic ground states remain robust across different degrees of chemical disorder. As summarized in Tables~\ref{tab:Mn_XA_L21}, starting from the fully-ordered $L2_1$-type structure, sequential swapping of Mn atoms in the 4$b$ sites with V atoms in the 8$c$ sites results in only minor reductions in the magnetic moment of V2 (nominal 4$d$ site), which then stabilizes as the structure approaches the $XA$-type configuration. Furthermore, despite atom swapping, the spin orientations at the corresponding Wyckoff sites are preserved. The magnetic moments of V atoms on both 4$c$ and 4$d$ sites decrease gradually from the $L2_1$ to the $XA$-type structure, as do the magnetic moments of Mn atoms on their respective sites. Overall, these results indicate that the intricate network of magnetic exchange couplings in V$_2X$Al is sufficiently robust to stabilize magnetic order even in the presence of substantial chemical disorder.

\begin{figure}
\begin{center}
\includegraphics[width=0.41\textwidth]{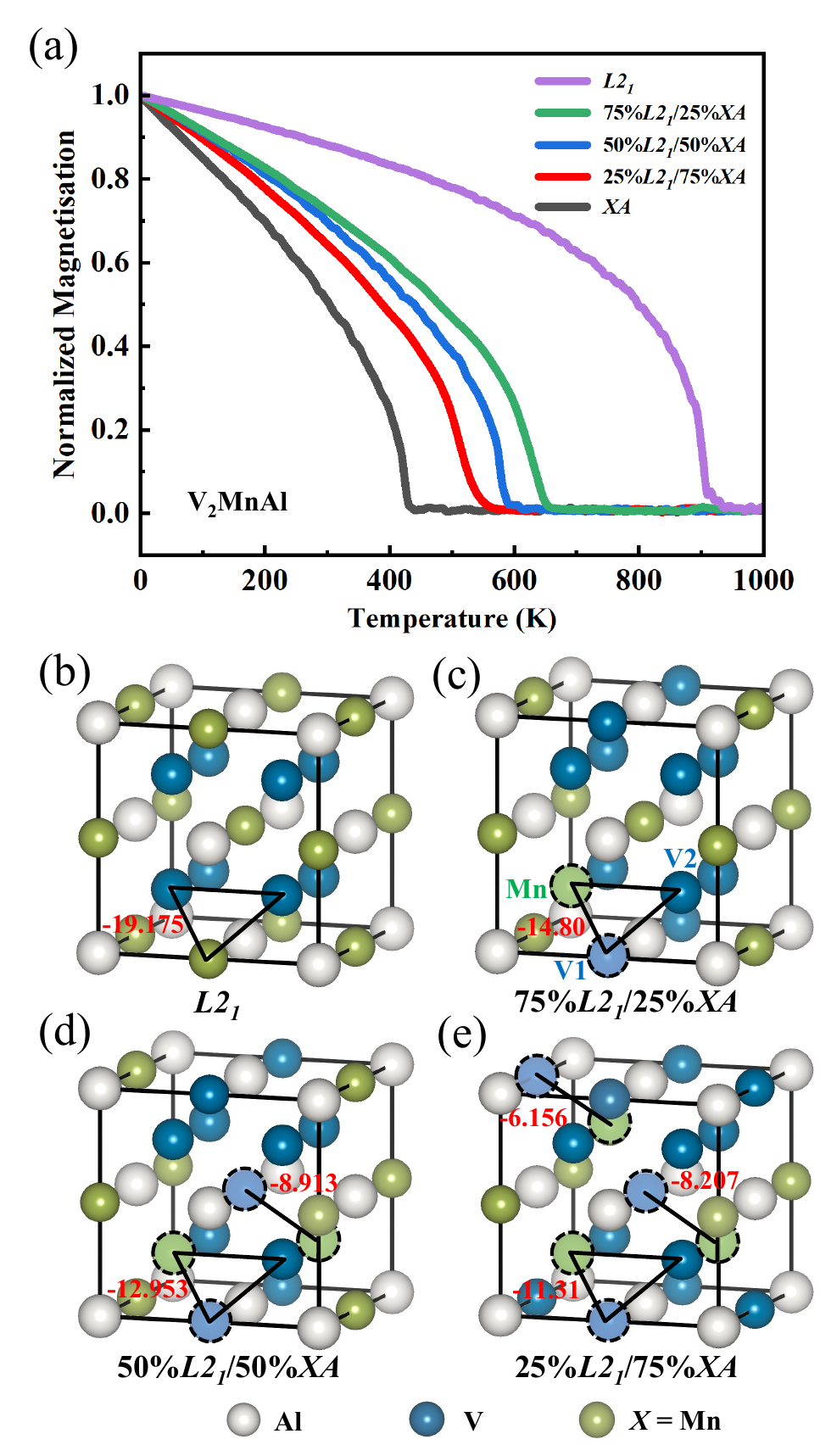}
\caption{(a) Temperature dependence of the normalized magnetic moment in partially-ordered V$_2$MnAl. (b–e) NN magnetic exchange couplings $J_{\mathrm{V-}X}$ for (b) fully-ordered $L2_1$, (c) 75\% $L2_1$, (d) 50\% $L2_1$, and (e) 25\% $L2_1$ partially-ordered structures.}
\label{fig5}
\end{center}
\end{figure}

The magnetic transition temperature in partially-ordered V$_2$MnAl gradually decreases as the structure evolves from $L2_1$ to $XA$-type structure. As shown in Fig.~\ref{fig5}(a), the Curie temperature $T_\mathrm{C}$ decreases from 900 K in the fully-ordered $L2_1$ structure to 420 K in the $XA$ configuration. Within our framework, the NN magnetic exchange couplings $J_{\mathrm{V-}X}$ and $J_{X-X}$ are the primary factors governing the magnetic transition temperature. In the partially-ordered structures studied here, swapping V and $X$ atoms directly affects the NN coupling $J_{\mathrm{V1-}X}$, while other magnetic exchange couplings are also modified. A comparison between the calculated Curie temperatures and the corresponding $J_{\mathrm{V1-}X}$ values in Fig.~\ref{fig5} shows a clear correlation. The magnitude of $J_{\mathrm{V1-}X}$ on the swapped sites closely tracks the variation of $T_\mathrm{C}$. For instance, in 75\% $L2_1$-V$_2$MnAl, $J_{\mathrm{V1-Mn}}$ is reduced to $-14.8$ meV, and further decreases to $-11.3$ meV ($-8.2$ meV and $-6.1$ meV for the other two relevant $X$ sites) in 25\% $L2_1$-V$_2$MnAl, compared to $-19$ meV in the fully-ordered $L2_1$ structure. The rapid reduction of $J_{\mathrm{V1-Mn}}$ from $L2_1$ to 75\% $L2_1$ is reflected in the corresponding decrease of $T_\mathrm{C}$ from 990 K to 650 K.

\begin{figure}
 \begin{center}
 \includegraphics[width=0.45\textwidth]{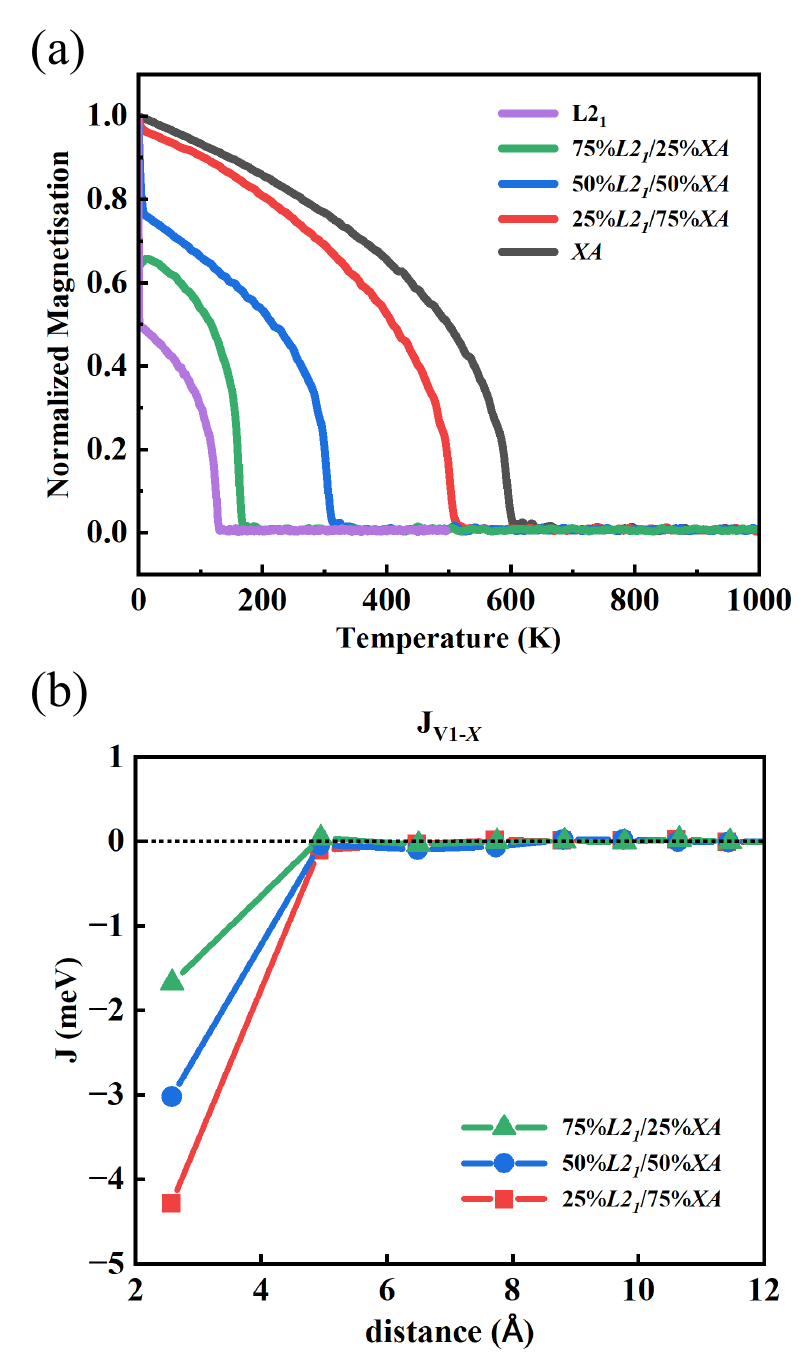}
 \caption{(a) Temperature-dependent normalized magnetic moments for partially-ordered V$_2$FeAl alloys. (b) Nearest-neighbor magnetic exchange couplings $J_{\mathrm{V1-}X}$ for the corresponding partially-ordered V$_2$FeAl structures.}
\label{fig6}
\end{center}
\end{figure}

Interestingly, within the V$_2$$X$Al series, the trend of magnetic transition temperatures in V$_2$FeAl is opposite to that observed in V$_2$MnAl. As shown in Fig.~\ref{fig6}(a), the Curie temperature $T_\mathrm{C}$ of V$_2$FeAl gradually decreases from 600 K to 120 K as the structure evolves from $XA$- to $L2_1$ type. This observation raises the question of which factors primarily govern this anomalous sequence of magnetic transition temperatures. As discussed above, the NN magnetic exchange coupling between V1 and $X$ ($J_{\mathrm{V1-}X}$) plays a dominant role in determining $T_\mathrm{C}$. Our calculations indicate that, although the sequence of $T_\mathrm{C}$ with increasing chemical order in V$_2$FeAl is reversed relative to V$_2$MnAl, the NN coupling $J_{\mathrm{V1-}X}$ remains a key factor. As shown in Fig.~\ref{fig6}(b), the absolute value of $J_{\mathrm{V1-}X}$ decreases systematically from the $XA$- to the $L2_1$-type structure, consistent with the observed reduction in $T_\mathrm{C}$. These results confirm the validity of $J_{\mathrm{V1-}X}$ as the primary determinant of the Curie temperature in partially-ordered V$_2$FeAl.

Our calculations on partially chemically disordered V$_2X$Al (for $X=$ Cr, Co, Ni, see Supplementary Information) indicate that the NN magnetic exchange coupling $J_{\mathrm{V1-}X}$ remains the dominant factor governing the magnetic transition temperature. Notably, during the transition from the $L2_1$ to the $XA$-type structure, the 4$d$ site is consistently occupied by V2 atoms, ensuring that the V-X-V magnetic triangular pathway is preserved. Consequently, the microscopic magnetic mechanism in chemically disordered V$_2X$Al is still primarily governed by the V-X-V triangular pathway and the associated NN magnetic exchange couplings $J_{\mathrm{V-}X}$.

\subsection{Electronic structures of  fully-ordered V$_2$$X$Al}\label{sec:elecstr}

\begin{figure}
 \begin{center}
 \includegraphics[width=0.48\textwidth]{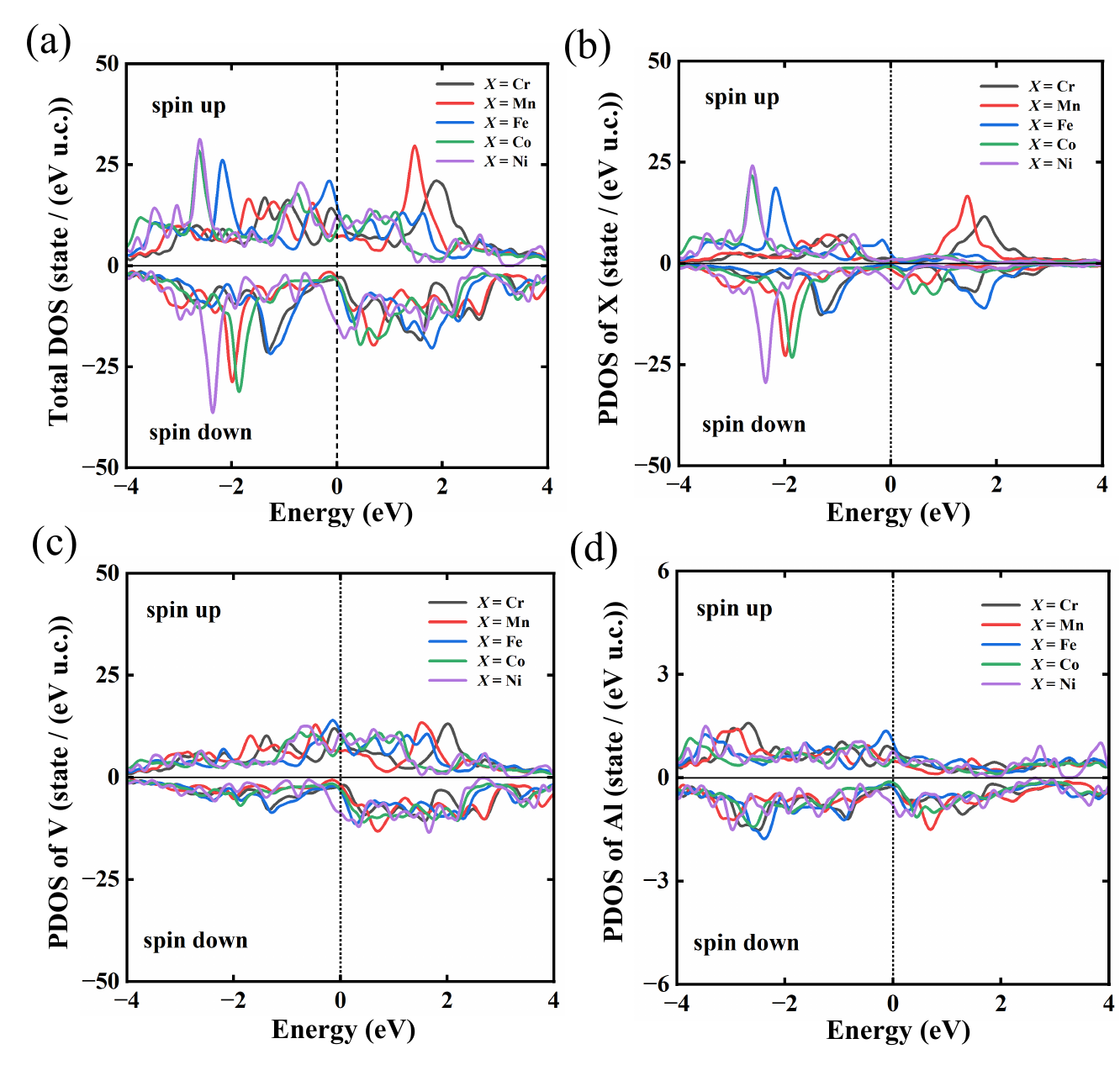}
 \caption{(a) Total density of states for $L2_1$-type V$_2X$Al ($X$ = Cr, Mn, Fe, Co, Ni). (b) Partial DOS projected onto the $X$ atoms in $L2_1$-type V$_2X$Al. (c) Partial DOS projected onto the V atoms in $L2_1$-type V$_2X$Al. (d) Partial DOS projected onto the Al atoms in $L2_1$-type V$_2X$Al.}
\label{fig7}
\end{center}
\end{figure}

Based on the calculated magnetic ground states of $L2_1$-type V$_2X$Al in Sec.~\ref{sec:magneticorder}, the observed complex magnetic orders and non-monotonic variation of total magnetic moments with atomic composition are proposed related to their electronic structures. In $L2_1$-type V$_2X$Al, the relatively small total magnetic moments of V$_2$CrAl and V$_2$MnAl originate from the anti-parallel alignment of spin moments between the V and $X$ sublattices. This behavior is clearly reflected in the calculated density of states (DOS). As shown in Fig.~\ref{fig7}(a), the total DOS of $L2_1$-type V$_2$$X$Al exhibits a high value at the Fermi level, indicating itinerant magnetism in these compounds \cite{Xu2024,Xu2025,Xu20252}. Notably, several compounds display nearly half-metallic character. As shown in Fig.~\ref{fig7}(b), Cr and Mn atoms exhibit pronounced occupied states with sharp peaks at approximately -1 eV and -2 eV in the spin-down channel, while their spin-up channel states are minimally occupied. In contrast, for $X=$ Fe, Co, and Ni, the PDOS of $X$ atoms shows significant occupation in the spin-up channel and comparatively low values in the spin-down channel. These electronic features are consistent with the observed transition in magnetic order from ferrimagnetism ($X=$ Cr, Mn) to ferromagnetism ($X=$ Fe, Co, Ni).

Furthermore, from $X=$ Fe to Ni, the PDOS of $X$ atoms shifts to lower energies in both spin-up and spin-down channels, as shown in Fig.~\ref{fig7}(b). This trend is attributed to the increasing number of valence electrons with rising atomic number. The pronounced difference between the spin-up and spin-down PDOS of Fe atoms indicates a large magnetic moment on Fe in V$_2$FeAl, whereas the PDOS of Ni atoms in V$_2$NiAl is nearly symmetric between spin channels, reflecting a significantly reduced magnetic moment. Examination of the V-atom PDOS from $X=$ Fe to Ni reveals an increase in spin-up states below the Fermi level for V atoms in Co- and Ni-containing compounds, particularly in the energy range from -1 to 0 eV, as shown in Fig.~\ref{fig7}(c), while the spin-down PDOS decreases in the same energy range. This behavior correlates with the observed increase in magnetic moments of V atoms from $X=$ Fe to Ni. In contrast, the PDOS of Al atoms remains largely unchanged across the series in Fig.~\ref{fig7}(d), consistent with the negligible magnetic moment on Al atoms. Taken together, the change in magnetic order from $X=$ Mn to Fe, combined with the opposing trends in magnetic moments of V and $X$ atoms from $X=$ Fe to Ni, accounts for the non-monotonic variation in the total magnetic moment across the V$_2$$X$Al series.

From our calculated electronic structures, we observe substantial hybridization among the V-$d$, $X$-$d$, and Al-$p$ states near the Fermi level in all compounds, as evidenced by the similar peak shapes and their alignment in energy. Although such hybridization is an evident feature throughout the V$_2X$Al series, the key factor governing the evolution of magnetic behavior is the variation in band filling associated with the $X$-site element. A comparison of the DOS for $X=$ Cr and Mn with those for $X=$ Fe, Co, and Ni clearly shows that, in the Cr and Mn systems, the $d$-electron manifold lies much closer to half-filling. This near half-filled configuration enhances antiferromagnetic superexchange between V and Cr (Mn) atoms through virtual electron-hopping processes between neighboring sites \cite{Sasioglu2008,Priolkar2013}. In contrast, as the $X$-site element becomes more electron-rich (from Fe to Ni), the increased band filling shifts the DOS away from half-filling, thereby suppressing the antiferromagnetic superexchange channel and promoting ferromagnetic exchange.

\section{CONCLUSION}

In this work, we employed density functional theory calculations combined with Monte Carlo simulations based on Heisenberg magnetic exchange couplings to investigate V$_2$$X$Al alloys and introduced the concept of ``magnetic motifs''. These magnetic motifs govern both the magnetic ground states and the magnetic transition temperatures in fully- and partially-ordered compounds. The underlying physical mechanism arises from the modulation of NN magnetic exchange couplings $J_{\mathrm{V-}X}$ within the V-$X$-V triangular pathways, which is influenced by atomic composition, crystal structure, and the degree of chemical disorder. Using the framework of magnetic motifs, the magnetic orders can be rationalized in terms of the V-$X$-V triangular pathways formed by NN $J_{\mathrm{V-}X}$ couplings. Furthermore, the magnetic transition temperatures are predominantly determined by both NN $J_{\mathrm{V-}X}$ and additional $J_{X-X}$ magnetic exchange interactions, irrespective of whether the alloy is fully or partially ordered. The concept of ``magnetic motifs'' captures the minimal set of dominant exchange interactions that governs both the magnetic ground states and the evolution of the Curie temperature across different structures and compositions. We anticipate that identifying such magnetic motifs in other Heusler alloys will provide a systematic and predictive framework for understanding and tailoring their diverse magnetic properties. Furthermore, a comprehensive investigation that combines phase stability analysis with detailed magnetic characterization will be essential for guiding the design of high-performance Heusler-based spintronic devices in the future studies.

\begin{acknowledgments}
This work is supported by the National Natural Science Foundation of China (Grants No. 52371174 and No. 12204033),  Science Challenge Project (Grant No.TZ2025009), the State Key Lab of Advanced Metals and Materials (Grant No. 2025Z-Z22). Numerical computations were performed at the High-performance Computing Platform of University of Science and Technology Beijing.
\end{acknowledgments}

\bibliography{ref}

\end{document}